\documentclass[twocolumn,showpacs,preprintnumbers,amsmath,amssymb]{revtex4}
\usepackage{tipa}
\usepackage{graphicx}
\usepackage{dcolumn}
\usepackage{bm}

\begin{document}

\title{$D$ and $D_s$ in mass loaded flux tube}
\author{Hong-Yun Shan$^1$\footnote{shywinner@shu.edu.cn}
and Ailin Zhang$^{1,2}$\footnote{Corresponding author:
zhangal@staff.shu.edu.cn}} \affiliation{$^1$Department of Physics,
Shanghai University, Shanghai, 200444, China\\
$^2$Kavli Institute for Theoretical Physics China, CAS, Beijing
100190, China}


\begin{abstract}
Heavy-light hadrons are studied in a mass loaded flux tube model.
The study indicates that the dynamics of mesons and baryons
containing a $c$ quark is well described by the mass loaded flux
tube. The hypothesis of good diquark-antiquark degeneracy is found
reasonable in heavy-light quark systems. The spectrum of charmed
($D$) and charmed strange ($D_s$) mesons is systematically computed.
$D$ and $D_s$ in $1D$ multiplets are predicted to have lower masses
in comparison with other theoretical predictions. The predicted
masses of the $1^-(1^3D_1)$ and the $3^-(1^3D_3)$ $D_s$ agree well
with those of recently observed $D_{s1}(2700)^{\pm}$ and
$D_{sJ}(2860)$, respectively.
\end{abstract}

\maketitle

\section{Introduction}
Hadrons spectrum can reveal the properties of the quark dynamics
such as the color confinement. So far, a great progress has been
made in the lattice QCD theory, but the quark dynamics in hadrons is
not very clear and the hadrons spectrum can not be extracted from
the QCD theory directly. The prediction of the hadrons masses has to
be made in all kinds of models, and an accurate prediction would be
a great challenge in hadrons spectroscopy. For heavy-light mesons,
the spectrum has been systematically computed in the relativized
quark model~\cite{gi}, heavy quark symmetry theory~\cite{ehq},
relativistic quark model~\cite{egf}, lattice QCD~\cite{lattice},
chiral quark model~\cite{pe} and some other
models~\cite{beh,br1,br2,br3,eeg}. In these calculations, it is
often difficult to predict the masses of higher orbital excited
states. In many cases, the predicted masses of the higher orbital
excited states seems to be overestimated in comparison with
experimental data.

In hadrons containing more than two quarks or antiquarks, two quarks
or antiquarks may attract each other to make a diquark or
anti-diquark cluster. The concept of diquark was put forth and was
extensively studied in strong
interactions~\cite{gell,di1,di2,di3,di4,di5,di6,di7}.

In terms of the diquark, a semi-classical mass loaded flux tube
model~\cite{sw} was recently exploited. In the model, a meson is
considered a system with a massive quark $m_1$ and a massive
anti-quark $m_2$ connected by a flux tube (or relativistic string)
with universal constant tension $T$ rotating with angular momentum
$L$. Similarly, a baryon is considered a system with a massive quark
$m_1$ and a massive diquark $m_2$ connected by the flux tube. The
flux tube is responsible for the color confinement. The mesons and
the baryons are therefore described by the same dynamics in the same
way. In addition, it is supposed that there is an approximate
degeneracy between a good diquark in baryons and a relevant
antiquark in mesons.

Light mesons and baryons have been studied and classified well in
the model~\cite{sw}. After the energy $E$ and the angular momentum
$L$ of the system are written down with the dynamical parameters in
the model, the $E$ can be expressed in the $L$ through some
reductions. The general form of $E$ is complicated, but the form is
simple in some special cases.

For the light quark systems, an approximate mass formula is
given~\cite{sw}
\begin{eqnarray}\label{eq1}
E\approx \sqrt{\sigma L}+\kappa L^{-{1\over 4}}\mu^{3\over 2},
\end{eqnarray}
where $T={\sigma \over 2\pi}$ is the string tension,
$\kappa\equiv\frac{2}{3}\frac{\pi^\frac{1}{2}}{\sigma^\frac{1}{4}}$,
and $\mu^{3\over 2}\equiv m^{3\over 2}_1+m^{3\over 2}_2$ with $m_1$
and $m_2$ are quark and antiquark/diquark masses.

The parameters for the light mesons and baryons in Eq.~(\ref{eq1})
were extracted from systematical analyzes of existing
data~\cite{sw,pdg04}. The analyzes indicated that the parameters for
the light mesons match well with the parameters for the light
baryons (see Table 7 in Ref.~\cite{sw}). The dynamics (especially
for large $L$) of light quark systems is well described by the mass
loaded flux tube~\cite{sw}. In the meantime, in order to account for
an approximate degeneracy between the $\Lambda$ baryons and the
relevant mesons, the hypothesis of "good diquark-antiquark
degeneracy" was proposed~\cite{sw}.

In heavy-light quark systems' case, an approximate mass formula was
reduced also~\cite{sw}
\begin{eqnarray}\label{eq2}
E=M+\sqrt{\frac{\sigma L}{2}}+2^{\frac{1}{4}}\kappa
L^{-\frac{1}{4}}m^\frac{3}{2},
\end{eqnarray}
where $M$ is the heavy quark mass and $m$ is the light quark/diquark
mass, other parameters are indicated in Eq.~(\ref{eq1}). The
spin-orbit forces are ignored and $L\neq 0$ in Eq.~(\ref{eq1}) and
Eq.~(\ref{eq2}).

However, the heavy-light quark systems have not been systematically
analyzed except that some $\Lambda_c$ baryons ($\Lambda_c(2285)$,
$\Lambda_c(2625)$, $\Lambda_c(2880)$) were simply mentioned(with the
relevant parameters $M_c=1600$ MeV, $m_{[ud]}=180$ MeV and
$\sigma=0.974$ GeV$^2$) in Ref.~\cite{sw}.

Many topics in the mass loaded flux tube model have not been studied
in the heavy-light quark systems. Whether the dynamics of the mesons
and the baryons can be described well by the flux tube has not been
examined. The spectrum of charmed and charmed strange mesons has not
been obtained. The hypothesis of "good diquark-antiquark
degeneracy", which holds in light quark systems, has not been
tested.

In this article, the heavy-light quark systems are studied in the
mass loaded flux model with the inclusion of the spin-orbit
interactions. The spectrum of the mesons containing one heavy $c$
quark/antiquark is systematically computed, and some possible
interpretations of recently observed states are discussed.

\section{Charmed and charmed strange mesons}

In the conventional quark model, mesons may be marked by their
quantum numbers $n^{2S+1}L_J$, where $n$ is the principle quantum
number, $S$ is the total spin, $L$ is the orbital angular momentum,
and $J$ is the total angular momentum. In most quark models, the
interactions between the quark and the antiquark include the
spin-independent confinement interaction, the spin-dependent
interactions (spin-orbit interaction, color hyperfine interaction)
and some other interactions~\cite{rgg,cornell,gi,bcps}. The
spin-orbit interaction consists of a color-magnetic piece and a
Thomas-precession piece. The spin-orbit interaction is often
considered the dominant one except for the confinement interaction,
and is sometimes simplified an $\vec L\cdot \vec S$ coupling. This
kind of spin-orbit interaction is employed in our study, while other
spin-dependent interactions such as the spin-spin interaction will
be ignored.

If the spin-orbit interaction was added to the energy $E$ of the
system from the beginning, the final relation between the $E$ and
the $L$ would be much more complicated than Eq.~(\ref{eq2}). As a
good approximation, a term $a\vec L\cdot \vec S$ responsible for the
spin-orbit interaction can be brought into Eq.~(\ref{eq2})
phenomenologically. The parameter $a$ is assumed a constant for the
mesons having the same flavors($a$ depends mainly on the heavy
flavor). It can be determined from the fit of experimental data. The
study in this article indicates that the mass loaded flux tube with
the inclusion of the $a\vec L\cdot \vec S$ coupling is very
potential to produce the whole $D$ and $D_s$ spectrum comparable to
the experimental data.

As well known, the heavy quark symmetry applies in the heavy-light
mesons. In the heavy quark limit, the mass and spin $s_Q$ of the
heavy quark decouples. All the mesons properties are determined by
the light degrees of freedom. The spin-parity $j^P$ (total angular
momentum $j=s_{\bar q}+l$ of light degrees of freedom) are good
quantum numbers and are conserved in strong interactions. In heavy
quark effective theory (HQET), the spin-dependent interactions
depend on $j$. A natural way to account for the spin-orbit
interaction in HQET is to include the $a\vec l\cdot \vec s_{\bar q}$
coupling instead of the $a\vec L\cdot \vec S$ coupling. However,
from our analysis of the experimental data, the $D$ mesons spectrum
is difficult to be reproduced with the simple inclusion of $a\vec
l\cdot \vec s_{\bar q}$ in Eq.~(\ref{eq2}). Besides, there may exist
a spin-orbit inversion problem~\cite{isgur} in HQET. The heavy quark
symmetry seems a little difficult to be accommodated in the present
flux tube picture. This difficulty is left as an open question and
is not studied here.

The reason of the inclusion of the $a\vec L\cdot \vec S$ coupling
can be realized in another way. The inclusion of the $a\vec L\cdot
\vec S$ in Eq.~(\ref{eq2}) will result in a nought of hyperfine
splitting (spin-triplet and spin-singlet splitting) of P-wave or
D-wave multiplet, which is consistent with theories and experiments.
These hyperfine splitting relations have already been predicted in
many quark potential models. The hyperfine splitting relations hold
very well in P-wave or D-wave multiplets of charmonium (even in $1P$
multiplets of $D$ mesons)~\cite{pdg08}.

Therefore, it is reasonable to extend the mass formula of the
heavy-light quark systems to
\begin{eqnarray}\label{eq3}
E=M+\sqrt{{\sigma L\over 2}}+2^{1\over 4}\kappa L^{-{1\over
4}}m^{3\over 2}+a \vec {L}\cdot \vec{S}
\end{eqnarray}
with
\begin{eqnarray*}
\vec L\cdot \vec S =\frac{J(J+1)-L(L+1)-S(S+1)}{2}.
\end{eqnarray*}

With this formula in hand, we go ahead with the study of the
heavy-light mesons. Firstly, we examine whether the hypothesis of
"good diquark-antiquark degeneracy" is favored. For this purpose,
the parameters $M_c=1600$ MeV and $\sigma=0.974$ GeV$^2$ extracted
from the charmed baryons~\cite{sw} are used as our inputs to compute
the spectrum of the $D$ mesons. Under the hypothesis of good
diquark-antiquark degeneracy, $m_{u,d}=m_{[ud]}=180$ MeV. We
obtained $m(1^1P_1)=2.406$ GeV for one $1P$ $D$ meson. This
predicted mass agrees well with that of the experimentally observed
$D_1(2430)^0$~\cite{pdg08}. Spectrum of other charmed mesons ($L>0$)
can be subsequently computed after $a=24.6$ MeV has been fitted from
other three $1P$ $D$ triplets. In terms of these parameters,
$m_s=320$ MeV is determined from $D_{s1}(2536)^\pm$ and
$D_{s2}(2573)^\pm$. The spectrum of $D_s$ mesons can be
systematically computed(the results are not given here for the
reason mentioned below).

The experimental spectrum of $D$ and $D_s$ mesons can be well
reproduced by the same group of parameters from the charmed baryons
except that the predicted $1^3P_0$ and $1^1P_1$ $D_s$ mesons are
much heavier in comparison with possible experimental candidates.
The fact that the spectrum of the $D$, $D_s$ mesons and the charmed
baryons is successfully obtained by the same formula and parameters
indicates explicitly that the dynamics of the mesons and the baryons
containing one heavy $c$ quark is well described by the flux tube.
It is found that the hypothesis of "good diquark-antiquark
degeneracy" is favored in the heavy-light quark systems.

In Ref.~\cite{sw}, the $\sigma$s are a little different for
different kinds of light mesons and baryons, for which there are two
reasons. One reason is that the $\sigma$s in the reference were
extracted with spin-orbit interactions ignored. The other reason is
that the string tension(the string is responsible for the dynamics)
may be different for hadrons containing different flavors.
Therefore, the $\sigma$s for mesons may be different from the
$\sigma$s  for baryons. In order to compute the spectrum of $D$ and
$D_s$ in a more reasonable way, the parameters of $\sigma$ and $a$
have to be refitted from the confirmed $D$ mesons.

For a consistent study, the masses of the $c$ quark and the light
$u,d$ quarks are regarded universal parameters for the mesons and
the baryons in our fitting processes. That is to say, the parameters
$m_c=1.6$ GeV and $m_{u,d}=180$ MeV extracted from the $\Lambda_c$
baryons are used as inputs to predict the spectrum of the charmed
mesons. Other parameters $\sigma=1.10$ GeV$^2$ and $a=37.9$ MeV are
extracted from the four $1P$ charmed mesons candidates(to extract
this two parameters, the minimum of mean square error of the mass of
the four $1P$ charmed mesons is applied). In terms of these
parameters, it is straightforward to get the spectrum of $1D$ and
$1F$ charmed mesons from Eq.~(\ref{eq3}).

In experimental side, each observed state has a mass uncertainty.
The mass uncertainties of observed states may result in some
uncertainties to our predictions. If a mass uncertainty $\pm 30$ MeV
is assumed for each $1P$ charmed candidate, the $\sigma$ will have
an uncertainty $\pm 0.09$ GeV$^2$. This assumed uncertainty may
result in $\pm 30$ MeV, $\pm 44$ MeV and $\pm 54$ MeV uncertainties
to the masses of the $1P$, $1D$ and $1F$ charmed and charmed strange
multiplets, respectively.

Our results for the charmed mesons are obtained in Table 1. In the
table, possible candidates for the $D$ mesons of each state are
displayed. For some states, there is no one to one correspondence
between the $j^P$ and the $n^{2S+1}L_J$ notation. To compare our
results with other theoretical predictions explicitly, we listed the
results of two typical computations~\cite{gi,pe}. In Ref.~\cite{gi},
the notation $n^{2S+1}L_J$ was used, and this notation is employed
in our calculation. The calculation was performed in HQET and the
notation $j^P$ was used in Ref.\cite{pe}. For simplicity, quantum
numbers $J^P$, $j^P$ and $n^{2S+1}L_J$ are all labeled in the table.
A parenthesis is put for the $j^P$ when there is no one to one
correspondence between the $j^P$ and the $n^{2S+1}L_J$ notation. A
dash "-" is put in the entry where the corresponding state has not
been computed in the two models. A "$?$" indicates that there is no
observed candidate corresponding to the assignment at present time.

Our results for the $1P$ states are comparable with those in
Refs.~\cite{gi,pe} and experiments. For the $1D$ states, our results
are much lower in comparison with those in Refs.~\cite{gi,pe}. This
obvious difference may provide a way to examine whether the mass
loaded flux tube model is reasonable. It may give people a hint to
find an underlying dynamics of hadrons.

\begin{table}
\begin{tabular}{lllllll}
Candidates~\cite{pdg08} & $J^P$ & $j^P$ &  $n^{2S+1}L_J$ &
GI~\cite{gi} & PE~\cite{pe}
& our paper\\
\hline\hline $D^0$ & $0^-$ & ${1\over 2}^-$  & $1^1S_0$ & $1.88$ & $1.868$ & - \\
$D^\star(2007)^0$ & $1^-$ & ${1\over 2}^-$ & $1^3S_1$ & $2.04$ & $2.005$ & -\\
\hline\hline $D^\star_0(2400)^0$ & $0^+$ & ${1\over 2}^+$ & $1^3P_0$ & $2.40$ & $2.377$ & $2.370$\\
$D_1(2420)^0$ & $1^+$ & $({3\over 2}^+)$ & $1^3P_1$ & $2.49$ & $2.417$ & $2.408$\\
$D_1(2430)^0$ & $1^+$ & $({1\over 2}^+)$ & $1^1P_1$ & 2.44 & 2.49 & 2.446\\
$D^\star_2(2460)^0$ & $2^+$ & ${3\over 2}^+$ & $1^3P_2$ & 2.50 & 2.46 & 2.484\\
\hline\hline ? & $1^-$ & ${3\over 2}^-$ & $1^3D_1$ & 2.82 & 2.795 & 2.623\\
? & $2^-$ & $({5\over 2}^-)$ & $1^3D_2$ & - & 2.775 & 2.699\\
? & $2^-$ & $({3\over 2}^-)$ & $1^1D_2$ & - & 2.833 & 2.737\\
? & $3^-$ & ${5\over 2}^-$ & $1^3D_3$ & 2.83 & 2.799 & 2.813\\
\hline\hline ? & $2^+$ & ${5\over 2}^+$ & $1^3F_2$ & - & 3.101 & 2.812\\
? & $3^+$ & $({7\over 2}^+)$ & $1^3F_3$ & - & 3.074 & 2.926\\
? & $3^+$ & $({5\over 2}^+)$ & $1^1F_3$ & - & 3.123 & 2.964\\
? & $4^+$ & ${7\over 2}^+$ & $1^3F_4$ & 3.11 & 3.091 & 3.078\\
\hline\hline
\end{tabular}
\caption{Spectrum of $D$ mesons(GeV) with parameters $\sigma=1.10$
GeV$^2$, $m_c=1.6$ GeV, $m_{u,d}=180$ MeV and $a=37.9$ MeV.}
\label{table-1}
\end{table}

In terms of the parameters $\sigma=1.10$ GeV$^2$, $m_c=1.6$ GeV and
$a=37.9$ MeV extracted from the charmed mesons and baryons, the
strange quark mass $m_s=288$ MeV is determined from two $1P$ charmed
strange mesons: $D_{s1}(2536)^\pm$ and $D_{s2}(2573)^\pm$. The
spectrum of the $D_s$ is subsequently computed and listed in Table
2.

$D^\star_{s0}(2317)^\pm$ and $D_{s1}(2460)^\pm$ are two "exotic"
states. They were firstly observed by BaBar~\cite{Babar1,pdg08} and
CLEO~\cite{cleo,pdg08} and were once interpreted as the $0^+~{1\over
2}^+$ and the $1^+~{1\over 2}^+$ $D_s$ mesons, respectively.
However, there are different interpretations to them. One difficulty
of the $D_s$ mesons interpretation is that they have lower masses in
comparison with theoretically predicted masses. So far, this two
states have not yet been pinned down definitely. In our article,
they are not used as inputs to determine the mass of the strange
quark. The difficulty of the $D_s$ mesons interpretation is not yet
solved in the mass loaded flux tube. $D^\star_{s0}(2317)^\pm$ and
$D_{s1}(2460)^\pm$ are really difficult to be interpreted as the
pure $1^3P_0$ and $1~P_1$($1~^3P_1$ will mix with $1~^1P_1$) $D_s$
mesons.

The situation of the $D_s$ mesons is similar to that of the $D$
mesons. The predicted masses of the $1D$ $D_s$ are much lower than
those in Refs.~\cite{gi,pe}.

\begin{table}
\begin{tabular}{lllllll}
 Candidates~\cite{pdg08} & $J^P$ & $j^P$ &  $n^{2S+1}L_J$ & GI~\cite{gi} & PE~\cite{pe}
& our paper\\
\hline\hline $D^\pm_s(1969)$ & $0^-$ & ${1\over 2}^-$ & $1^1S_0$ & 1.98 & 1.965 & - \\
$D^{\star\pm}_s(2112)^0$ & $1^-$ & ${1\over 2}^-$ & $1^3S_1$ & 2.13 & 2.113 & -\\
\hline\hline $D^\star_{s0}(2317)^\pm$ & $0^+$ & ${1\over 2}^+$ & $1^3P_0$ & 2.48 & 2.487 & 2.478\\
$D_{s1}(2536)^\pm$ & $1^+$ & $({3\over 2}^+)$ & $1^3P_1$ & 2.57 & 2.535 & 2.516\\
$D_{s1}(2460)^\pm$ & $1^+$ & $({1\over 2}^+)$ & $1^1P_1$ & 2.53 & 2.605 & 2.554\\
$D_{s2}(2573)^\pm$ & $2^+$ & ${3\over 2}^+$ & $1^3P_2$ & 2.59 & 2.581 & 2.592\\
\hline\hline $D_{s1}(2700)^{\pm}$ & $1^-$ & ${3\over 2}^-$ & $1^3D_1$ & 2.90 & 2.913 & 2.714\\
? & $2^-$ & $({5\over 2}^-)$ & $1^3D_2$ & - & 2.900 & 2.789\\
? & $2^-$ & $({3\over 2}^-)$ & $1^1D_2$ & - & 2.953 & 2.827\\
$D_{sJ}(2860)$ & $3^-$ & ${5\over 2}^-$ & $1^3D_3$ & 2.92 & 2.925 & 2.903\\
\hline\hline ? & $2^+$ & ${5\over 2}^+$ & $1^3F_2$ & - & 3.224 & 2.894\\
? & $3^+$ & $({7\over 2}^+)$ & $1^3F_3$ & - & 3.247 & 3.008\\
? & $3^+$ & $({5\over 2}^+)$ & $1^1F_3$ & - & 3.203 & 3.046\\
? & $4^+$ & ${7\over 2}^+$ & $1^3F_4$ & 3.19 & 3.220 & 3.160\\
\hline\hline
\end{tabular}
\caption{Spectrum of $D_s$ mesons(GeV) with parameters $\sigma=1.10$
GeV$^2$, $m_c=1.6$ GeV, $m_s=288$ MeV and $a=37.9$ MeV.}
\label{table-2}
\end{table}

Recently, two new $D_s$ candidates were observed. $D_{sJ}(2860)$ was
first reported by BaBar~\cite{BaBar} in
\begin{eqnarray*}
D_{sJ}(2860)\to D^0K^+~,~D^+K^0_s
\end{eqnarray*}
with $M=2856.6\pm 1.5(stat)\pm 5.0(syst)$ and $\Gamma=48\pm
7(stat)\pm 10(syst)$ MeV. For its natural spin-parity:
$J^P=0^0,~1^-,~\cdots$, this state was explained as the first radial
excitation of the $D^\star_{s0}(2317)$ or the
$3^-(1^3D_3)$~\cite{br,cfn,ctls,zhu}.

$X$(2690) was also reported by BaBar~\cite{BaBar}, but the
significance of the signal was not stated.

$D_{sJ}$(2700) was first observed by Belle~\cite{belle1} in
\begin{eqnarray*}
B^+\to \bar D^0D_{sJ}\to\bar D^0D^0K^+
\end{eqnarray*}
with $M=2715\pm 11^{+11}_{-14}$ and $\Gamma=115\pm 20^{36}_{-32}$
MeV. The mass and the decay width change a little in their published
version~\cite{belle2}. For its $J^P=1^-$, this state was interpreted
as a mixture of the $2^3S_1$ and the $1^3D_1$~\cite{ctls} or the
$1^-(1^3D_1)$~\cite{zhu}.

In these interpretations, one difficulty for the $1^3D_1$ and the
$1^3D_3$ $D_s$ interpretations is that the masses of
$D_{s1}(2700)^{\pm}$ and $D_{sJ}(2860)$ are $100\to 200$ MeV lower
than the theoretical predictions. In the mass loaded flux tube
model(Table 2), there is no difficulty to these interpretations at
all. The predicted mass of the $1^3D_1$ $D_s$ is around $2714\pm 30$
MeV and the predicted mass of the $1^3D_3$ $D_s$ is around $2903\pm
44$ MeV. As the masses and the decays modes considered, it is very
possible that $D_{s1}(2700)^{\pm}$ and $D_{sJ}(2860)$ are the
$1^3D_1$ and the $1^3D_3$ charmed strange mesons, respectively.


\section{Conclusions}

In summary, the mass loaded flux tube with the inclusion of the
spin-orbit interaction is studied. In heavy-light quark systems, the
mesons and the baryons are well described by the mass loaded flux.
Experimental data (spectrum) for the masses of the mesons and the
baryons containing one heavy $c$ quark are reproduced well by the
same formula and the same parameters. It is found that the
hypothesis of "good diquark-antiquark degeneracy" is a reasonable
and consistent hypothesis in heavy-light quark systems.

Our results indicate that $D^\star_{s0}(2317)^\pm$ and
$D_{s1}(2460)^\pm$ are unlike the pure $1^3P_0$ and $1~P_1$ charmed
strange mesons, respectively.

Our predictions of masses of the $1D$ $D$ and $D_s$ are much lower
in comparison with other theoretical predictions. The predicted
masses of the $1^-(1^3D_1)$ and the $3^-(1^3D_3)$ charmed strange
mesons agree well with those of the recently observed
$D_{s1}(2700)^{\pm}$ and $D_{sJ}(2860)$ states, respectively. Other
two $1D$ charmed strange mesons around $2800$ MeV are expected to
exist, which is left for the confirmation of future experiments.

Of course, many observed states are mixed states in the real world.
Under mixing, how to interpret the observed states with the pure
states is not clear, which deserves more study.

The heavy-light quark systems containing one $b$ quark have not been
analyzed. The heavy quarkonium has not been explored either. The
systems with radial excitation or excitation inside the string are
not yet involved. How to extend the model to compute the spectrum of
all kinds of mesons and baryons would be an interesting work.
Furthermore, how to develop the model to describe the production and
decay dynamics deserves further exploration. It will be more
important to find whether there is an underlying dynamics in the
mass loaded flux tube model different from existing QCD inspired
models.

The mass loaded flux tube model is a semi-classical one, it will be
interesting to study the mass loaded flux tube in a "fundamental"
theory such as the string theory(some features such as the Regge
trajectory behavior in the mass loaded flux tube model have already
been obtained in the string theory).

Acknowledgments: This work is supported by the National Natural
Science Foundation of China under the grant: 10775093.

\end{document}